\documentclass[aps,pre,epsf,superscriptaddress,amsmath,amssymb,amsfonts,twocolumn,showpacs]{revtex4-1}

\usepackage{graphicx}
\usepackage{epsfig}
\usepackage{dcolumn}
\usepackage{bm}
\usepackage{braket}
\usepackage{amsmath}
\usepackage{color}
\newcommand{\abs}[1]{\left| #1 \right|} 
\usepackage[utf8]{inputenc}
\usepackage[left=2cm,right=2cm,top=0.0cm,bottom=2cm,includeheadfoot]{geometry} 

\usepackage{hyperref}
\usepackage{multirow}
\DeclareMathOperator{\sech}{sech} 
\usepackage{todonotes}
\extrafloats{100}

\begin{document}
\title{Phase Diagram, Stability and Magnetic Properties of Nonlinear Excitations \\ in Spinor Bose-Einstein Condensates}

\author{G. C. Katsimiga}
\affiliation{Center for Optical Quantum Technologies, Department of Physics,
University of Hamburg, Luruper Chaussee 149, 22761 Hamburg,
Germany}

\author{S. I. Mistakidis}
\affiliation{Center for Optical Quantum Technologies, Department of Physics,
University of Hamburg, Luruper Chaussee 149, 22761 Hamburg,
Germany}

\author{P. Schmelcher}
\affiliation{Center for Optical Quantum Technologies, Department of Physics,
University of Hamburg, Luruper Chaussee 149, 22761 Hamburg,
Germany} \affiliation{The Hamburg Centre for Ultrafast Imaging,
University of Hamburg, Luruper Chaussee 149, 22761 Hamburg,
Germany}

\author{P. G. Kevrekidis}
\affiliation{Department of Mathematics and Statistics, University
of Massachusetts Amherst, Amherst, MA 01003-4515, USA}

\date{\today}

\begin{abstract}
We present the phase diagram, the underlying stability and magnetic properties as well as the 
dynamics of nonlinear solitary wave excitations arising in the distinct phases of a harmonically confined spinor 
$F=1$ Bose-Einstein condensate. 
Particularly, it is found that nonlinear excitations in the form of dark-dark-bright  
solitons exist in the antiferromagnetic and in the easy-axis phase of a 
spinor gas, being generally unstable in the former while possessing stability intervals in the latter phase.
Dark-bright-bright solitons can be realized in the polar and 
the easy-plane phases as unstable and stable configurations respectively; the latter phase can also feature 
stable dark-dark-dark solitons.
Importantly, the persistence of these types of states upon transitioning, by means of tuning the 
quadratic Zeeman coefficient from one phase to the other is unravelled. Additionally, the 
spin-mixing dynamics of stable and unstable matter waves is analyzed, revealing among others the 
coherent evolution of magnetic dark-bright, nematic dark-bright-bright and dark-dark-dark 
solitons.
Moreover, for the unstable cases unmagnetized or magnetic droplet-like 
configurations and spin-waves consisting of regular and magnetic solitons are seen to dynamically 
emerge remaining thereafter robust while propagating for extremely large evolution times. 
Interestingly, exposing spinorial solitons to finite temperatures, 
their anti-damping in trap oscillation is showcased. It is found that 
the latter is suppressed for stronger bright soliton component ``fillings''.  
Our investigations pave the wave for a systematic production and analysis involving spin transfer 
processes of such waveforms which have been recently realized in ultracold experiments. 
\end{abstract}

\maketitle

\section{Introduction}

Ultracold atoms constitute ideal platforms for investigating the nonlinear behavior of quantum many-
body systems due to their high degree of controllability and isolation from the 
environment~\cite{Pethick,Stringari,Bloch}.
A principal example has been the exploration of dark and bright
solitons and their dynamical manifestations, as well as their
multi-dimensional and multi-component extensions
in Bose-Einstein condensates (BECs)~\cite{abdullaev,siambook}.  
Indeed, a variety and admixtures of these types of excitations are nowadays known to exist 
in scalar~\cite{Burger,Denschlag,Khaykovich,Weller,Frantz_dark,Lamporesi}, 
pseudo-spinor~\cite{Busch,Becker,Hamner,Yan,Garrett,Prinari} 
and spinor~\cite{Ieda1,Ieda2,Uchiyama,Alejandro,Lannig} BECs.
Importantly, in recent years, many of these works have featured
experimental realizations of such excitations.
However, up to now the majority of both theoretical and experimental endeavors has been mainly 
focused on studying solitons in single and pseudo-spinor BEC systems
and also within the so-called Manakov limit~\cite{Manakov}. 
The latter assumes the intra- and the inter-species coupling to be on equal footing.
As such the physics of nonlinear excitations outside this limit is 
less explored although there is an ongoing theoretical effort in this
direction over the past few years~\cite{Karamatskos,Tsitoura,Katsimiga_inter,Katsimiga_bif_coll,Hannes}.
It is also relevant to note in passing that in the quantum regime and under suitable
conditions, coherent structures such as bright solitons have been found to be promising candidates for
quantum metrology beyond the Heisenberg limit~\cite{Tsarev}.

Arguably, even less explored appears to be the connection between regular e.g. 
vector solitons and magnetic solitons or higher spin objects such as $F=1$ spinors and spin-waves.
Namely, nonlinear structures for which the magnetic interactions between the species are 
a crucial component. 
For instance, a magnetic soliton typically residing in a spin 
balanced density background~\cite{QuPitaString} is characterized by a localized spin magnetization, measured as 
the difference between the population of the participating components.
Such nonlinear polarization waves have also been studied earlier~\cite{Kamchatnov}
in binary BECs for parametric variations lying outside the Manakov limit
both in the absence and in the presence of a Rabi coupling between the ensuing 
components~\cite{QuTylStringPita}.
Case examples of magnetic solitons are dark and dark-bright (DB) matter waves that differ from 
their regular or standard counterparts in a two-fold manner: (i) they exist for unequal intra and 
interspin couplings and (ii) their width scales according to the spin-healing length~\cite{QuPitaString}.
They can also have the form of dark-antidark solitons, with the latter being density bumps on top 
of the BEC background, that have been very recently experimentally 
monitored~\cite{Danaila,Farolfi,Chai,Us_DADs}.

$F=1$ spinor BECs offer the possibility for studying not only regular solitons but also magnetic 
ones and admixtures thereof.
In particular, owing to the far richer phase diagram exhibited 
by such gases~\cite{Kawaguchi} (see, also,~\cite{Oberthaler2020} for a
recent discussion and~\cite{Simos} for the impact of many-body
effects) already several works have been devoted in studying a variety of nonlinear excitations that arise 
in them~\cite{Boris_MI,Dabrowska,Zhang1,Ektor,Szankowski,Ueda}.
These include for instance spin domains~\cite{Miesner,Swislocki}, spin textures~\cite{Ohmi,Song}, 
the very recently experimentally observed dark-dark-bright (DDB) and dark-bright-bright (DBB) 
solitons~\cite{Bersano} (and variants~\cite{Meng,Liu}, as well as
interactions~\cite{Lannig} thereof) and even twisted magnetic 
solitons~\cite{Ueda_Twisted}. 

However, and despite the fact that a fair amount of previous works have been devoted to studying 
the nonlinear excitations that arise in this spinor context, several important questions still remain open.
A major one concerns the principal phase diagram of existence (and stability) of solitonic excitations. 
Yet another interesting perspective, in view also of the intense
ongoing  interest on magnetic spinor solitons~\cite{Farolfi,Chai},   
is the classification of such excitations in terms of their magnetic properties. 
Furthermore, the persistence (and stability) of such entities and their morphing under 
a quadratic Zeeman (QZ) energy shift variation are currently vastly unexplored. 
Accordingly, the coexistence of distinct solitonic configurations in the same phase and the fate of the latter 
in the presence of finite temperature are far less appreciated. 
In the present work we exploit the substantial ongoing momentum spearheaded 
by recent experiments~\cite{Bersano,Farolfi,Chai,Lannig,raman2} 
and address the aforementioned open aspects.

In particular, we first map out the complete phase diagram of nonlinear 
excitations arising in one-dimensional (1D) harmonically confined spinor $F=1$ BECs when accounting for both 
antiferro- and ferro-magnetic spin-dependent interactions. 
This phase diagram, which to the best of our knowledge has never been extracted thus far, is subsequently 
explored in detail, including the connection to the stability of the emergent waveforms. 
More precisely, DDB, DBB and dark-dark-dark (DDD) solitons constitute its principal ingredients. 
DDB solutions exist in the antiferromagnetic (AF) and the easy-axis (EA) phase, DBB solitons 
arise in the polar (PO) and the easy-plane (EP) phases and DDD waves are realized in the EP phase too.

Moreover, we unveil the largely unexplored magnetic properties not only of the principal spinor solitons 
emerging in each phase of the system, but also of their ensuing deformations for varying QZ energy shifts. 
An exhaustive study of the stability properties of the involved in each phase soliton solutions is offered, 
along with the relevant outcome when crossing, in terms of a QZ energy shift variation, the distinct phase transition thresholds. 
The latter facilitates a fruitful direction for near future experimental realizations dealing with such metastable states. 
Interestingly, the dynamical evolution of stable and unstable configurations 
(whose longevity suggests their experimental relevance) reveals among others: the coherent 
evolution of magnetic DB solitons and spin-mixing processes leading to changes in the magnetic properties 
of the evolved entities including the formation of composite spin objects. 
The latter are composed of regular solitons and spin-waves. 
Additionally we observe metastable states evolving into periodically recurring unmagnetized 
Thomas-Fermi (TF)-droplet configurations and also magnetized entities with droplets occupying the symmetric spin 
sublevels --with a domain wall (DW) separating them-- and a localized wavefunction hosted in the remaining spin-component. 
The latter nearly periodic structures closely resemble magnon drops~\cite{Macia,Divinskiy}, 
while in both cases DWs are imprinted in the local magnetization. 
The above composite dynamically generated spin configurations were unprecedented thus far. 
Finally, the fate of spinor solitons at finite temperatures is explored, unveiling their anti-damped 
(growing amplitude) in-trap oscillation. 
The latter, is found to be suppressed for stronger bright soliton component ``fillings" of the dark notch 
generalizing this way earlier findings regarding single~\cite{PGK_darkFT} and two-component~\cite{PGK_DBFT} 
BECs to the spin-1 setting.  

Our work is structured as follows. In section~\ref{theory} the relevant mean-field theoretical framework is introduced. 
The ground state (GS) phase diagram of a harmonically trapped 1D spin-1 BEC is initially discussed in Section~\ref{phase_diagram} 
and we then proceed to the presentation and systematic exploration of the relevant phase diagram of nonlinear excitations 
in the form of DDD, DDB and DBB solitons. 
Section~\ref{stability} addresses the existence, the stability properties, 
by means of Bogoliubov de-Gennes (BdG) linearization analysis, and subsequently the dynamics of 
the different solitonic waveforms that arise in the distinct phases of the spinor system. 
Finally, in section~\ref{conclusions} we summarize our findings and also provide future perspectives.

\section{Spinor Setup and Magnetization Measures}\label{theory} 

A spin-1 BEC composed of the magnetic sublevels $m_F=0,\pm 1$ of the hyperfine state $F=1$, 
either of a $^{87}$Rb~\cite{Bersano} or a $^{23}$Na~\cite{Stenger} atom gas being 
confined in a 1D harmonic trap is considered.
A cigar-shaped geometry is employed that has been very recently realized 
experimentally~\cite{Bersano} utilizing a highly anisotropic trap with the longitudinal 
and transverse trapping frequencies obeying  
$\omega_x \ll \omega_{\perp}$. 
In the mean-field framework the dynamics of such a spinor system can be described 
by the following coupled dimensionless Gross-Pitaevskii equations (GPEs) 
of motion~\cite{Martikainen,Dabrowska,Ektor,Xiong}
\begin{eqnarray}
i \partial_t \Psi_{0} &=&
\mathcal{H}_{0}\Psi_{0}  
+ c_{0} \left( |\Psi_{+1}|^2 + |\Psi_{0}|^2 + |\Psi_{-1}|^2 \right)\Psi_{0}\nonumber\\
&+&c_{1}\left( |\Psi_{+1}|^2 + |\Psi_{-1}|^2 \right)\Psi_{0}
+ 2c_{1}\Psi_{+1}\Psi^{*}_{0} \Psi_{-1}, \nonumber\\
\label{eq:pm1} 
\end{eqnarray}
for the $m_F=0$ magnetic sublevel, while the symmetric $m_F=\pm1$ spin-components obey
\begin{eqnarray}
i \partial_t \Psi_{\pm 1} &=&
\mathcal{H}_{0}\Psi_{\pm 1}  
+ c_{0} \left( |\Psi_{+1}|^2 + |\Psi_{0}|^2 + |\Psi_{-1}|^2 \right)\Psi_{\pm 1}\nonumber\\
&+&c_{1}\left( |\Psi_{\pm1}|^2 + |\Psi_{0}|^2 - |\Psi_{\mp 1}|^2 \right)\Psi_{\pm 1}+q\Psi_{\pm 1}\nonumber\\
&+&c_{1}\Psi^{*}_{\mp 1}\Psi^{2}_{0}.
\label{eq:0} 
\end{eqnarray}
In Eqs.~(\ref{eq:pm1})-(\ref{eq:0}), $\Psi_{m_F}(x,t)$ denotes the wavefunction 
of the $\ket{F=1, m_F=0}$ and $\ket{F=1, m_F=\pm1}$ spin-components respectively.
The single particle Hamiltonian term is $\mathcal{H}_0\equiv -\frac{1}{2} \partial_{x}^2 +V(x)$, 
with $V(x)=\frac{1}{2}\Omega^2 x^2$ being the 1D harmonic potential. 
Here, $\Omega\equiv\omega_{x}/\omega_{\perp}$
plays the role of the longitudinal over the transverse trapping frequency
and is typically a small parameter i.e., $\Omega \ll 1$~\cite{Pethick,Stringari}. 
Additionally, $q$ denotes the QZ energy shift parameter
that leads to an effective detuning of the $m_{F}=\pm 1$ spin-components with respect to the 
$m_{F}=0$ one.
It is quadratically proportional to an external magnetic field applied along the spin-$z$
direction~\cite{Kawaguchi,Zhang} 
and can be experimentally tuned by either adjusting the applied magnetic field~\cite{Santos} or by 
using a microwave dressing field~\cite{Leslie,Bookjans}. 

Moreover, $c_{0}$ and $c_{1}$ are the so-called spin-independent and spin-dependent interaction 
coefficients. 
The former accounts for attractive (repulsive) interatomic interactions upon 
taking negative (positive) values and the latter is positive ($c_{1}>0$) for antiferromagnetic 
and negative ($c_{1}<0$) for ferromagnetic interactions.
Both $c_{0}$ and $c_{1}$ are expressed in terms of the $s$-wave scattering lengths $a_{0}$ and 
$a_{2}$, accounting for two atoms in the scattering channels with total spin $F=0$ and $F=2$ 
respectively, via the relations $c_0=\frac{(a_0+2a_2)}{3a_\perp}$ and 
$c_1=\frac{(a_2-a_0)}{3a_\perp}$~\cite{Alejandro,Dabrowska}.  
Here, $a_\perp=\sqrt{\hbar/M\omega_{\perp}}$ is the transverse harmonic oscillator length
with $M$ denoting the mass, e.g., of a $^{87}$Rb atom. 
Eqs.~(\ref{eq:pm1})-(\ref{eq:0}) have been made dimensionless by measuring length, energy and time 
in units of $\sqrt{\hbar/ (M\omega_\perp)}$, $\hbar\omega_\perp$ and $\omega_\perp^{-1}$ 
respectively.
Consequently, the corresponding interaction strengths are expressed in 
terms of $\sqrt{\hbar^3\omega_\perp/M}$. 
In the adopted units and for a ferromagnetic, i.e., $c_1<0$, spinor BEC of ${}^{87}$Rb 
atoms~\cite{Oberthaler2020,Alejandro}, the experimentally measured spin-dependent 
and spin-independent couplings also used herein are 
$c_1 \approx -5\times 10^{-3}\sqrt{\hbar^3\omega_\perp/M}$ and 
$c_0=1\sqrt{\hbar^3\omega_\perp/M}$.

Additionally, the population of each spin-component is defined as 
\begin{eqnarray}
n_{m_F}&=&\frac{1}{N}\int dx |\Psi_{m_F}|^2, \hspace{0.3cm} m_{F}=0,\pm 1. 
\label{Particles}
\end{eqnarray}
Here, $N=\sum_{m_F}\int dx |\Psi_{m_F}|^2$ denotes the total number of particles that is a 
conserved quantity for the spinorial system of Eqs.~(\ref{eq:pm1})-(\ref{eq:0}).
Evidently, $0\leq n_{m_F}\leq 1$ is satisfied. 
Furthermore, in order to quantify first- and second-order transitions between the distinct phases
of the spin-1 BEC system as well as to monitor the magnetic properties of the emergent nonlinear 
excitations during evolution we utilize the magnetization along the
spin-$z$-axis that reads
\begin{eqnarray}
M_{z}&=&\frac{1}{N}\int dx \left(|\Psi_{+1}|^2-|\Psi_{-1}|^2\right). \label{Magn}
\end{eqnarray}
$M_{z}$ essentially measures the population imbalance between the symmetric $m_{F}=\pm1$ 
components and $-1\leq M_{z}\leq 1$. 
For instance, a fully magnetized state along the $+z$ or $-z$ spin direction corresponds 
to $M_{z}=+1$ or $M_{z}=-1$ respectively.
To encounter also possible population transfer between the $m_{F}=0$
and the $m_{F}=\pm1$ spin states we invoke the polarization of the spinorial setting defined 
as follows~\cite{Kawaguchi}
\begin{eqnarray}
P&=&\frac{1}{N}\int dx \left[|\Psi_{0}|^2-\left(|\Psi_{+1}|^2+|\Psi_{-1}|^2\right)\right].
\label{Pol}
\end{eqnarray}
It can be easily deduced that $-1\leq P\leq 1$.
As we will unveil later on, $P$ also accounts for alterations in the magnetic properties 
of the spinor system and allows us to distinguish among fully magnetized and unmagnetized
spin configurations as we cross, by means of varying the QZ energy shift $q$, a 
phase transition boundary.
\begin{figure*}[htb]
 \centering \includegraphics[width=0.88\textwidth]{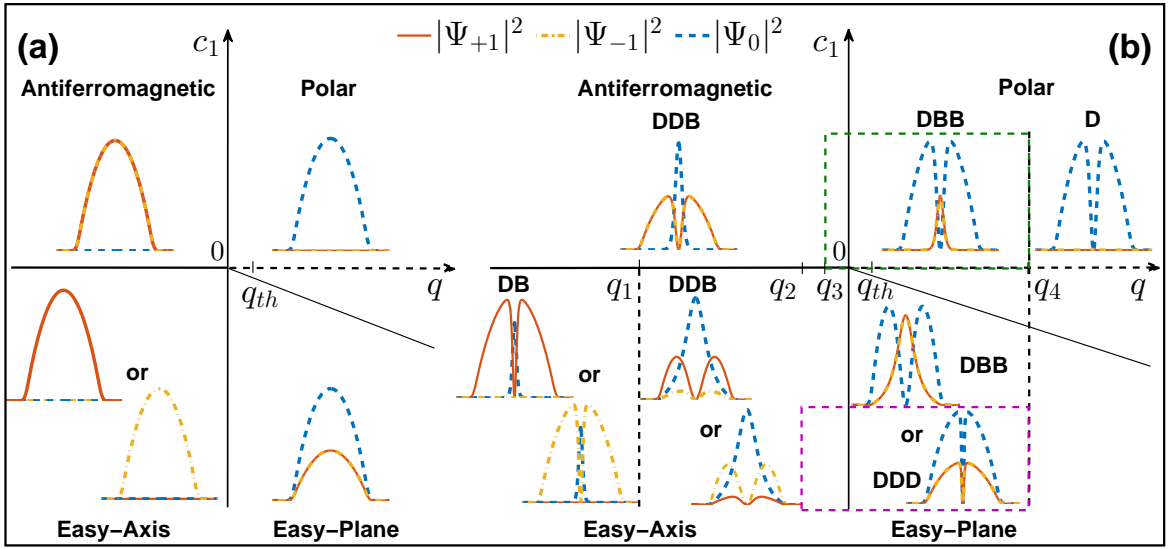}
 \caption{Schematic illustration of (a) the GS phase diagram of a harmonically trapped 
 spin-1 BEC in the $(c_1, q)$ plane.
 (b) The corresponding phase diagram of nonlinear excitations having the form of DDD, DDB and DBB 
 solitons. 
 In both cases characteristic density profiles, $|\Psi_{m_F}|^2$ with $m_{F}=0,\pm1$, 
 of each of the phases that can be realized in such a system are provided (see legend). 
 For $c_{1}>0$ the AF and the PO phases occur for $q<0$ and $q>0$ respectively.
 For $c_{1}<0$ the system is: in the EA ferromagnetic phase if $q<0$, in the EP phase if 
 $0<q<q_{th}\equiv 2n|c_1|\approx 0.017$ and in the PO one if $q\geq q_{th}$.  
 In both phase diagrams solid black lines mark the individual phase transition boundaries. 
 Vertical dashed black lines designate the regions from where and on the distinct configurations 
 deform, i.e., {the DDB into DB ($q_{1}\approx -0.515$) or the DBB to D ($q_{4}\approx 0.994$) 
 or similarly the DDD to D ($q_{4}\approx 0.016$) solitons.}
 Dashed purple (green) box indicates that DDD (DBB) solitons exist also within the EA (AF) phase 
 while their existence terminates at $q_{2}\approx -0.016$ ($q_{3}\approx -0.005$). 
 The specific threshold values refer to $\mu_{0,\pm1}=2$.
  \label{Figure1}}
\end{figure*}

Finally, for the numerical investigations that follow, 
the trapping frequency is fixed to $\Omega=0.1$, but we
note that the results presented herein are not altered even 
for trapping frequencies of the order of $\Omega=0.01$
that are used in recent spin-1 BEC experiments~\cite{Bersano}.
This way our findings can be experimentally realized e.g. by using a transversal confinement 
frequency $\omega_{\perp}=2\pi \times 175$Hz ($\omega_{\perp}=2\pi \times 380$Hz) 
along with a longitudinal one $\omega_{x}=2\pi \times 1.4$Hz~\cite{Bersano} ($\omega_{x}=2\pi \times 5.4$Hz~\cite{Chai}) 
for a $^{87}\rm{Rb}$ ($^{23}\rm{Na}$) spinor gas. 
For the above selection of frequencies, evolution times of the order of $t\sim 10^3$ typically 
monitored herein correspond to $t \approx 0.91$s ($t\approx 0.42$s) in dimensional units for 
a $^{87}\rm{Rb}$ ($^{23}\rm{Na}$) gas. 
Only slight deviations of the corresponding transition boundaries are observed. 
For instance, for the EP to PO transition, while $q_{th}\equiv 2n|c_1|=0.02$ 
(with $n$ denoting the peak density) for $\Omega=0.01$ it is $q_{th}\approx 0.017$ for $\Omega=0.1$. 
Additionally, $c_{0}=1$, $c_{1}=\pm 5\times 10^{-3}$ and we choose the chemical potentials 
of the different components $\mu_{0,\pm1}=2$. 
It is also important to mention that we have checked that the results  to be presented below are 
robust also for $c_{0}=1$ and $c_{1}=3.6\times 10^{-2}$, namely for the experimentally relevant 
interaction coefficient parameter ratio corresponding to ${^{23}}$Na gas and also for larger chemical potentials, 
i.e. $\mu_{0,\pm1}=3$ and $\mu_{0,\pm1}=5$. 
To access the distinct phases of the spinor system, we typically vary $q$ within the 
intervals $[-1.5, 0.5]$ and $[-0.5, 1.5]$. 
Moreover, in order to identify the existence of stationary states a fixed-point numerical iteration scheme, 
based on Newton's method, is employed~\cite{NewtonKrylov}. 
To simulate the dynamical evolution of the distinct DDD, DDB and DBB solitons 
governed by Eqs.~(\ref{eq:pm1})-(\ref{eq:0}), a fourth-order Runge-Kutta integrator is utilized 
while a second-order finite differences method is used for the spatial derivatives. 
The spatial and time discretization are $dx = 0.05$ and $dt = 0.001$ respectively. 
Our numerical computations are restricted to a finite region by employing hard-wall boundary 
conditions. 
Particularly, in the dimensionless units adopted herein, the hard-walls are located 
at $x_{\pm}=\pm 80$ and we do not observe any appreciable density for $\abs{x} > 20$.

\section{Phase Diagram of nonlinear Excitations}\label{phase_diagram}

Before delving into the details of the phase diagram of nonlinear excitations in the form of DDD, 
DDB and DBB solitons that arise in spin-1 BECs, we first briefly revisit the relevant GS 
phase diagram of a harmonically confined 1D spin-1 BEC~\cite{Oberthaler2020}.  
This description will enable us to qualitatively expose the effect of embedding nonlinear 
structures into the different magnetic phases. 

\subsection{Ground state phase diagram}
A schematic representation of the GS phase diagram 
is illustrated in Fig.~\ref{Figure1}(a).
As it has been recently demonstrated~\cite{Kawaguchi,Oberthaler2020,Simos}
different phases can be realized for such a confined spin-1 system.
They stem from the interplay between the sign of the spin-dependent 
interaction coefficient $c_1$ and the strength of the QZ term $q$.
Specifically, for $c_1>0$, $q<0$ the system is in the AF phase
with equally populated $m_F=\pm 1$ spin-components thus having an unmagnetized GS 
[see (Eq.~\ref{Magn})]. 
The latter is indeed characterized by $M_z=0$ and $P=-1$. 
A first order phase transition~\cite{Carr,Sachdev} separates this phase from the PO
one that can be reached upon increasing $q$. 
The transition point appears at $q=0$ and the PO phase is characterized again by an 
unmagnetized GS but with all atoms populating the $m_F=0$ spin-component.  
Therefore $M_z=0$ and $P=1$.
On the other hand, for $c_1<0$, $q<0$ the system resides in the EA phase.
Its GS is fully magnetized either along the $+z$ or the $-z$ spin-direction,
i.e. either the $m_F=+1$ or $m_F=-1$ spin state is populated. 
As a result $M_z=+1$ or $M_z=-1$ respectively and $P=-1$. 
Upon increasing $q$ a second-order phase transition occurs at $q=0$ 
and for $0<q<q_{th}\approx 0.017$
the system enters the EP phase with its GS having all three $m_F$ 
components populated. 
Particularly here, $M_{z}=0$ reflecting the fact that the $m_F=\pm 1$ spin states are equally 
populated while $P \in \left(-1, 1\right)$.  
Finally, for $q>q_{th}\approx 0.017$ yet another second order phase transition  
takes place which leads to an unmagnetized GS having only the $m_F=0$ spin 
component populated, i.e. $M_z=0$ and $P=1$. In this case, once more the PO phase is reached.

Note here, that in order to obtain the above-discussed GS phase diagram
TF profiles are employed as initial guesses, within our fixed point 
algorithm~\cite{NewtonKrylov}, for the distinct $m_F=0,\pm 1$ states having the form
\begin{eqnarray}
\Psi_{m_F}(x,t=0)=\sqrt{c^{-1}_0\big(\mu_{m_F}-V(x)\big)}.
\label{TF}
\end{eqnarray}
When this expression is used here and below, it is implied to be valid when the quantity under the radical 
is non-negative (and the relevant wavefunction is padded with zeros outside that region).
In Eq.~(\ref{TF}), $\mu_{m_F}$ denotes the chemical potential of each spin-component while
a stationary state satisfies the phase matching condition $\mu_{0}=\left(\mu_{+1}+\mu_{-1}\right)/2$~\cite{Ektor_DW,Oberthaler2020}.
\begin{figure*}[htb]
 \centering \includegraphics[width=0.85\textwidth]{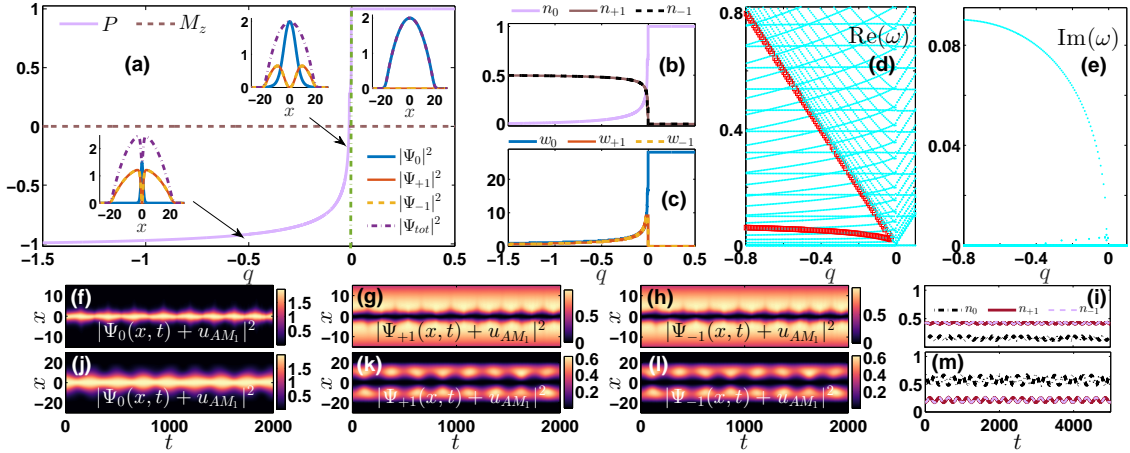}
  \caption{(a) Polarization $P$, and magnetization, $M_z$, of the spin-1 system for a DDB state 
 existing within the AF phase upon increasing the QZ coefficient $q$ in order to 
 enter the PO phase. Vertical dashed-dotted (green) line marks this transition.
 Insets from bottom left to top right illustrate characteristic soliton profiles for $q=-0.5$, 
 $q=-0.015$ and $q=0.1$ (see also black arrows). 
 (b) Population, $n_{m_{F}}$, and (c) soliton width, $w_{m_{F}}$ of the DDB solitons as a function of $q$ (see legends). 
(d), (e) BdG spectrum of stationary DDB solitons for an AF to PO transition, depicting respectively the real, ${\rm Re}(\omega)$, and 
the imaginary part, ${\rm Im}(\omega)$, of the involved eigenfrequencies, $\omega$, as $q$ is varied. 
The trajectories of the two anomalous modes (AMs) present in this spectrum are indicated by red squares (see text).  
The eigenfrequency zero crossings occur at $q=0$.   
 (f)-(h) [(j)-(l)] Dynamical evolution of the density, $|\Psi_{m_{F}}+u_{AM_1}|^2$ 
 of a DDB soliton being excited by the eigenvector, $u_{AM_1}$, associated
with the lowest-lying anomalous mode (AM) appearing in the aforementioned spectrum for $q=-0.1$ [$q=0$]. 
(i), (m) Temporal evolution of the populations, $n_{m_F}(t)$, for the 
 above selection of $q$'s. In all cases $m_{F}=0,\pm 1$ (see legends) while the remaining system parameters correspond to 
 $\Omega=0.1$, $\mu_{0,\pm1}=2$, $c_{1}=5\times 10^{-3}$, and $c_{0}=1$.
  \label{Figure2}} 
 \end{figure*}

\subsection{Phase diagram of solitonic excitations}
In order to unravel the phase diagram of nonlinear excitations depicted in Fig.~\ref{Figure1}(b),
the spin-1 system is initialized in each of the above-identified phases embedding 
dark and bright solitons as wavefunctions for each of the $m_{F}=0,\pm1$ states.
In particular, the standard, stationary solitonic waveforms used read~\cite{Bersano} 
\begin{eqnarray}
\Psi_D(x,t=0)&=& \sqrt{c^{-1}_{0}\left[\mu_{m_F}-V(x)\right]}\tanh\left(Dx\right), 
\label{dark}\\ 
\Psi_{B}(x,t=0)&=& \eta\sech\left(Dx \right). 
\label{bright}
\end{eqnarray}
In the above expressions $\Psi_D(x)$ and $\Psi_B(x)$ denote the 
wavefunctions utilized for a dark and a bright soliton configuration respectively. 
In Eq.~(\ref{dark}) the quantity under the square root denotes the customary used 
TF background needed for dark solitons to be embedded on.
Moreover, $D$ and $\eta$ refer, respectively, to the common inverse width considered for each 
spinorial soliton component and the amplitude 
of the bright soliton configuration (see our detailed discussion in Sec.~\ref{stability}).

It is found that DDB solitons, being unmagnetized configurations, 
exist within the AF phase for all values of the QZ energy shift
lying within the interval $q\in(-1.5, 0)$, with the dark solitons 
effectively trapping the bright one appearing in the $m_{F}=0$ spin-component.
This trapping mechanism becomes progressively less effective.
Namely, as $q$ increases towards the 
phase transition point ($q=0$) the bright soliton gradually 
becomes the dominant configuration before morphing into a TF one.
We remark that the existence of a DDB soliton in the AF phase already presents
fundamental deviations from its GS properties.
Indeed, in the latter case the $m_{F}=0$ magnetic sublevel is unpopulated (of course also the
$m_{F}=\pm 1$ states do not feature a dark soliton in the relevant GS).    
On the contrary, DBB solitons, being again unmagnetized configurations, are identified in the PO 
phase, namely for $q\in[0, 0.994)$ which deform towards a single dark soliton 
occupying the $m_{F}=0$ component for $q>0.994$.
Remarkably these states persist even upon decreasing $q$ so as to enter the AF phase 
until a critical value of the QZ energy shift, i.e. $q_{3}\approx -0.005$, is 
reached [see dashed green box in Fig.~\ref{Figure1}(b)]. 
Note that such DBB configurations also constitute excited states within the PO phase 
since for the GS only the $m_{F}=0$ state is occupied. 
Turning to $c_{1}<0$, stationary solutions of the DDB type are realized within the ferromagnetic EA 
phase existing within the  parametric region  $q\in(-0.515, 0.007)$. 
These DDB states deform as $q$ decreases further into fully magnetized, 
i.e. $M_{z}=+1$ ($M_{z}=-1$), DB solitons that occupy the $m_{F}=0$ and 
$m_{F}=+1$ ($m_{F}=0$ and $m_{F}=-1$) components.
Once again this is far from the GS of the EA featuring only  $m_{F}=+1$ (or $m_{F}=-1$) populations.    
Moving to $q>0$, namely entering the EP phase, two types of solitonic solutions are found to exist
for the spin-1 system.
These excitations can have the form of unmagnetized spinor DBB or DDD solitons,
a result that is permitted by the relevant GS where all three 
magnetic components are occupied.
The former solitonic entities appear to be significantly broader when compared to the more 
localized DDD configurations and become highly localized as we enter the PO phase. 
Recall that the PO GS supports population only in the $m_{F}=0$ magnetic sublevel.
The transition point for the DBB configuration appears at
$q_{4}\approx 0.994$ while it occurs significantly earlier, $q_{4}\approx 0.016$, for the DDD state. 
Decreasing the QZ term, $q$, in order to enter the EA phase reveals that the DBB configuration 
deforms fast, around $q\approx-0.007$, to a metastable state with two TF wavefunctions occupying 
the $m_F=\pm1$ components.
Contrary to this deformation, DDD solitons continue to exist within the EA phase for values up to 
$q_{2}\approx-0.016$ before their transitioning towards two darks that occupy the symmetric 
$m_F=\pm1$ components [see dashed purple box in Fig.~\ref{Figure1}(b)].

\section{Stability analysis and dynamics of spinor solitons}\label{stability}

Our aim in what follows is not only to illustrate the existence of stationary spinor solitons of the 
DDD, DDB and DBB type existing in a 1D harmonically confined spin-1 BEC composed 
e.g. of $^{87}$Rb atoms and obeying Eqs.~(\ref{eq:pm1})-(\ref{eq:0}), but also 
to systematically investigate their stability properties. 
We remark that for ferromagnetic (AF) BECs we consider $c_{1}=-5\times 10^{-3}$ ($c_{1}=5\times 10^{-3}$) 
as representative example and vary the QZ energy shift to access the underlying magnetic phases.  

\subsection{Antiferromagnetic DDB matter waves}

For instance, in order to infer about the existence of DDB solitons within the AF phase  
shown in the phase diagram of Fig.~\ref{Figure1}(b), the matter wave dark solitons
of Eq.~(\ref{dark}) are embedded as initial guesses for the $m_F=\pm 1$ spin-components and 
the bright soliton of Eq.~(\ref{bright}) is utilized for the $m_F=0$ spin state.
Employing the above ansatz, and using the iterative scheme
discussed above, DDB stationary states are found within the AF phase, 
i.e. for $c_{1}=5\times 10^{-3}$ and for values of $q \in (-1.5, 0)$.
Characteristic DDB density profiles, $|\Psi_{0,\pm1}|^2$, are presented as insets 
in Fig.~\ref{Figure2}(a).
However, upon increasing $q$ towards the transition point ($q=0$) above which the PO phase is 
realized, the DDB solitons deform into states where the bright
structure  in the $m_F=0$ component overfills/dominates the dark wells.
Also the total density, $|\Psi_{\rm{tot}}|^2$, of the spinor system exhibits a 
TF profile instead of the dark-shaped density appearing deep in the AF phase.
This altered nature of the DDB configuration, which remains unmagnetized ($M_{z}=0$) for all
values of $q$, is naturally accompanied by a change in the polarization of this configuration. 
The DDB solitons possess $P=-1$ for $q\leq-1.5$ reflecting the fact that deep in the AF phase 
only the $m_{F}=\pm 1$ components bearing dark solitons are populated [Fig.~\ref{Figure2}(b)], 
while the polarization takes values $-1<P\leq 0$ as we approach the transition point.
At $q\approx-0.02$ all three components are equally populated having
significantly wider~\cite{FWHM} stationary states  [Fig.~\ref{Figure2}(c)] as
compared to the ones for larger negative $q$ values.
This broadening suggests that the DDB character of the relevant states is lost.
Importantly, at $q=0$ an abrupt population transfer to the $m_F=0$ 
component [Fig.~\ref{Figure2}(b)] associated with the drastic deformation 
of this latter configuration to the GS of the PO phase 
manifests itself; see e.g. the right uppermost inset of Fig.~\ref{Figure2}(a).

In order to extract the stability properties of the aforementioned DDB stationary states 
(as well as for the DBB and DDD solitons to be presented below), 
a linear stability or BdG analysis is performed. 
The latter consists of perturbing the iteratively identified in each phase stationary solutions 
$\Psi^{0}_{m_F}(x)$ (with $m_{F}=0,\pm1 $) through the ansatz
\begin{eqnarray}
\Psi_{m_F}(x,t) &=& \Big[\Psi^{0}_{m_F}(x)+\epsilon \left(a_{m_F}(x)e^{-i \omega t} 
+b^{*}_{m_F}(x)e^{i \omega t}\right)\Big] \nonumber\\
&\times & e^{-i \mu_{m_F} t}.
\label{bdg}
\end{eqnarray}
By inserting this ansatz into the system of Eqs.~(\ref{eq:pm1})-(\ref{eq:0})
and linearizing with respect to the small amplitude parameter $\epsilon$ leads to 
an eigenvalue problem for the eigenfrequencies $\omega$, or 
equivalently eigenvalues $\lambda\equiv -i \omega$, and eigenfunctions 
$(a_{0},b_{0},a_{+1},b_{+1},a_{-1},b_{-1})^T$ that is solved numerically. 
For further details on the BdG analysis we refer the reader to 
Refs.~\cite{siambook,Kevre,Skryabin}. 
Due to the generally complex nature of the ensuing eigenfrequencies, it becomes apparent that the 
following possibilities can arise: if modes with purely real eigenvalues or equivalently imaginary 
eigenfrequencies or complex eigenvalues/eigenfrequencies are identified, these are responsible 
for the existence of an instability~\cite{Kevre}. 
The former case is referred to as an exponential instability, while the latter as an oscillatory 
instability, as the growth is non-monotonic, but rather involves oscillations.
\begin{figure*}[htb]
\centering \includegraphics[width=0.85\textwidth]{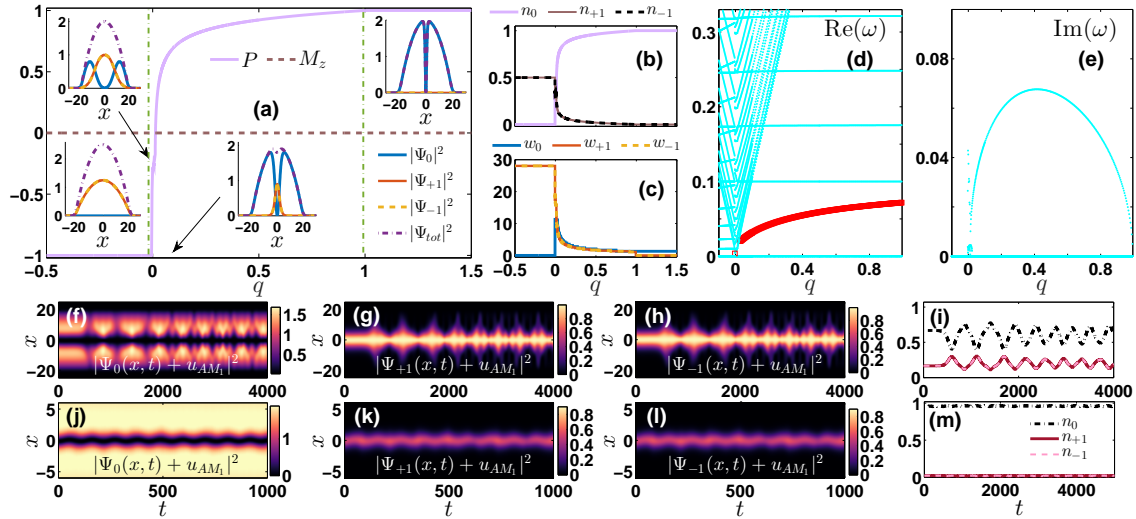}
\caption{(a) Polarization $P$, and magnetization, $M_z$, for DBB solutions existing within the 
 PO phase upon decreasing the QZ coefficient $q$ towards the AF phase. 
 Insets from bottom left and on illustrate characteristic soliton profiles for $q=-0.5$, 
 $q=-0.004$, $q=0.1$ and $q=1$ (see also black arrows). 
Vertical dashed-dotted (green) lines at $q\approx -0.005$ and $q\approx 0.994$ 
mark the boundaries of deformation of the DBB wave for different $q$'s. 
(b) Populations, $n_{m_{F}}$, and (c) soliton widths, $w_{m_{F}}$, for varying $q$. 
(d), (e) Stability analysis outcome showcasing respectively the real, ${\rm Re}(\omega)$, and  
the imaginary part, ${\rm Im}(\omega)$, of the relevant eigenfrequencies under a $q$ variation.
The trajectory of the single AM appearing in this spectrum is indicated by red squares (see text).
The destabilization of the DBB state occurs at $q_{cr}\approx-0.004$
while for $q\geq 1$ only a single dark exists in the $m_{F}=0$ component. 
(f)-(h) [(j)-(l)] Density evolution, $|\Psi_{m_{F}}+u_{AM_1}|^2$, of a perturbed PO DBB soliton
for $q=0.024$ [$q=0.5$].
(i), (m) Temporal evolution of the populations, $n_{m_F}(t)$, for the aforementioned values of $q$. 
In all cases $m_F=0,\pm1$ while $\Omega=0.1$, $\mu_{0,\pm1}=2$, $c_{1}=5\times 10^{-3}$ 
and $c_{0}=1$.
\label{Figure5}}
 \end{figure*}
Moreover, due to the Hamiltonian structure of the system investigated herein, 
quartets of such eigenfrequencies can occur~\cite{Skryabin}. Namely if $\omega$
is an eigenfrequency so are $-\omega$ and $\pm\omega^{*}$. 
As such, if $\rm{Im}(\omega)\neq 0$, then there will always exist a mode leading to the growth 
and eventual deformation of the examined in each phase solitonic configuration.

The BdG analysis outcome for the DDB soliton solutions is shown in Fig.~\ref{Figure2}(d), (e).   
DDB solitons constitute excited states of the spin-1 system, exactly like their two-component 
dark-bright analogue~\cite{Katsimiga_inter}, a feature that is reflected in their linearization 
spectra via the emergence of the so-called anomalous modes (AMs). 
These eigenstates are quantified via the negative energy or negative Krein 
signature~\cite{Skryabin} defined for the spinor system as 
\begin{eqnarray}
\begin{split}
  K= \Omega \int \Big(&|a_{0}|^2 - |b_{0}|^2 +|a_{+1}|^2\\ 
  &- |b_{+1}|^2+|a_{-1}|^2 - |b_{-1}|^2 \Big) dx.
  \label{krein}
  \end{split}
\end{eqnarray}
The existence of these modes is central to our stability analysis since their 
potential collision with positive Krein signature modes can give rise
to stability-changing events in the form of oscillatory instabilities
or Hamiltonian-Hopf bifurcations~\cite{Skryabin}. 
Such modes illustrate the feature that the solution is {\it not} a GS, but rather an excited state of the system. 
Indeed, when the relevant frequencies remain real (see also below), their negative energy suggests 
that while the waveform is stable dynamically, it is not stable thermodynamically~\cite{siambook}. 
Should then, a channel of energy dissipation be available 
(as, e.g., in the thermal condensates discussed below), then these eigendirections would lead to 
instability enabling the waveform to transition to the desired minimum energy state. 
However, there is an additional key role of negative energy modes which is crucial {even in the case of $T=0$ BECs}. 
More specifically, upon variation of parameters (like  $q$ and $c_1$ considered herein) these modes may 
collide with breathing modes of the condensate. 
This collision is also topologically necessitated (from the theory of AMs) to lead to a so-called 
oscillatory instability, which is featured via oscillatory (rather than purely exponential) growth. 
Hence, these AMs may be responsible for the manifestation of instabilities even in the zero temperature regime.
In the present analysis the AMs are denoted by red squares and the background ones with light blue dots. 
The DDB solution possesses, due to the presence of two dark solitons~\cite{Kapitula0}, two such modes 
[Fig.~\ref{Figure2}(d)] that cross the origin of the spectral plane at $q_{cr}=0$ signalling the 
destabilization of the DDB wave [Fig.~\ref{Figure2}(e)]. 
Interestingly, and also for all values of $q\in (-1.5,0)$, it is found that the eigenvector associated with 
the lowest-lying AM causes an overall shift when added to the stationary DDB solution. 
This in turn implies that a perturbed, with this eigenvector, DDB soliton will perform an oscillatory 
motion within the parabolic trap. 
On the contrary, the eigenvector corresponding to the higher-lying AM, besides a weak displacement, 
further leads to an asymmetric DDB configuration. 
This asymmetry, as we shall show later on, is responsible for the breathing motion of the 
DDB entity and its effect is dominant with respect to the aforementioned shift. 
It is this higher-lying AM that is responsible for the generic instability, i.e. the one with the larger 
imaginary contribution, shown in Fig.~\ref{Figure2}(e). 
As such, the remaining loop bifurcation illustrated in this figure can be directly assigned 
to the lowest among the two AMs depicted in Fig.~\ref{Figure2}(d).	

However, for values of $q$ closer to the critical point, defining the AF to PO transition 
boundary, the destabilization of both modes leads, irrespectively 
of which mode we excite (namely the first 
lower-lying one or the second), to a breathing motion of the DDB
configuration.
Its response is visualized in the spatio-temporal evolution of the densities, 
$|\Psi_{m_{F}}(x,t)+u_{AM_1}|^2$, presented in Fig.~\ref{Figure2}(f)-(h) entailing 
both the particle-like oscillation of the DDB soliton but predominantly the
overall breathing of the state.
Although the first AM is excited in this case, the 
dynamics is dominated by the breathing mode.
The nature of this composite motion is also reflected in the irregular oscillation
of the population, $n_{m_F}(t)$, of each $m_F$ component illustrated in Fig.~\ref{Figure2}(i).
At the transition/destabilization point the prevailing feature of the perturbed DDB configuration 
is its breathing as can be seen by monitoring the evolution of the densities illustrated in 
Fig.~\ref{Figure2}(j)-(l) together with the coherent oscillation 
of the relevant populations [Fig.~\ref{Figure2}(m)].
In both of the aforementioned cases the oscillatory character of $n_{m_F}(t)$
implies a weak amplitude spin-mixing dynamics.
\begin{figure*}[htb]
 \centering \includegraphics[width=0.95\textwidth]{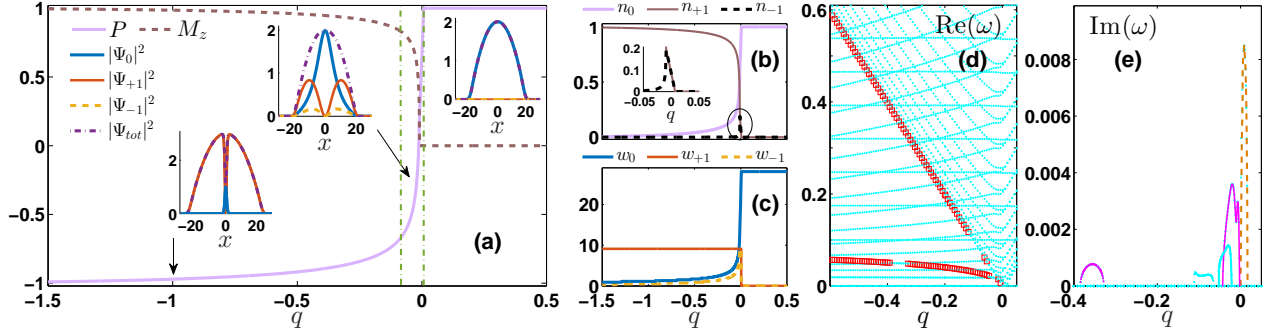}
 \caption{(a) Polarization, $P$, and magnetization, $M_z$, of the spinor system for a DDB 
 state existing within the EA phase upon varying the QZ coefficient $q$ so as to enter 
 the EP and PO phases. Vertical dashed-dotted (green) lines mark the boundaries of 
 deformation of the DDB as $q$ is varied (see text). Insets from bottom left to top right
 illustrate characteristic wave profiles for $q=-1.0$, $q=-0.011$ and $q=0.1$ (see also black arrows). (b) Populations, 
 $n_{m_{F}}$, and (c) soliton widths, $w_{m_{F}}$, as functions of $q$ with $m_F=0,\pm1$ (see legends).
 Inset in (b) provides a magnification of the region close to the corresponding transition point.
 (d), (e) Real, ${\rm Re}(\omega)$, and imaginary part, ${\rm Im}(\omega)$, of $\omega$ for varying $q$. 
 In (d) the trajectories of the two emergent AMs are indicated by red squares while in (e) 
 the distinct loop bifurcations marked by magenta and light blue lines corresponding respectively to the lower- and higher-lying AM 
 occur for $q \in$: $[-0.384, -0.329]$, $[-0.111, -0.066]$, $[-0.052, -0.022]$, $[-0.042, -0.003]$ 
 and $[0, 0.016]$. The dashed brown line is used to denote the destabilization of both AMs.  
 The rest of the parameters are $\Omega=0.1$, $\mu_{0,\pm1}=2$, 
 $c_{1}=-5\times 10^{-3}$, and $c_{0}=1$.
 \label{Figure9}}
 \end{figure*}

\subsection{Polar DBB solitons}

Next we turn to the PO phase which is characterized by $c_1=5\times 10^{-3}$ and $q>0$ 
[see also Fig.~\ref{Figure1}(a)].
According to the phase diagram of Fig.~\ref{Figure1}(b), here one can identify stationary DBB 
soliton solutions for values of $q\in [0, 0.994)$. 
Specifically and so as to capture the occurrence of these solitonic waveforms 
we utilize, within our fixed point iteration scheme, the dark soliton ansatz of Eq.~(\ref{dark})
as an initial guess for the $m_F=0$ spin-component, 
while bright solitons given by Eq.~(\ref{bright}) 
are considered for the remaining symmetric $m_{F}=\pm1$ components.    
These states are characterized by zero magnetization, preserving 
this way the magnetic properties of the GS within this phase, 
but they have a polarization that 
acquires values $-1<P<1$ [Fig.~\ref{Figure5}(a)]. 
In particular, $P=1$ for $q>1$, i.e. deep in the PO phase, and it gradually decreases 
as $q\rightarrow 0^{+}$ all the way to $P=-1$ for $q<0$.  
This latter behavior of $P$ reveals in turn that despite the fact that the GS 
configuration does not support all three $m_{F}$ components to be populated 
this is not the case for the respective nonlinear excitations [Fig.~\ref{Figure5}(b)].
Selected DBB soliton profiles are depicted as insets in Fig.~\ref{Figure5}(a).  
From these profiles it can be deduced that these unmagnetized DBB waves exist not only within the 
above-provided $q$ interval but also at ($q=0$) and below ($q\in [0, -0.005)$) 
the transition point that separates the PO and the AF phases. 
However, as we approach the transition point from above $q\rightarrow 0^{+}$
the DBB states deform towards wider configurations [Fig.~\ref{Figure5}(c)] 
featuring a pronounced bright soliton component that dominates. 
This dominant bright component results in turn to a $|\Psi_{\rm{tot}}|^2$ that has a TF profile 
instead of a $\tanh$-shaped one occurring for values of $q$ well inside the PO phase.
For $q<-0.005$ an abrupt transition leads to a metastable configuration in which 
the $m_F=\pm1$ are equally populated having also minimal polarization ($P=-1$), 
see the bottom left inset of Fig.~\ref{Figure5}(a).   
On the contrary, for $q>0.994$ yet another but gradual this time deformation of the DBB matter waves
towards a dark soliton with maximal polarization ($P=+1$) occupying the $m_F=0$ spin 
state occurs [top right inset of Fig.~\ref{Figure5}(a)]. 

By investigating the stability properties of the above solitonic entities, it is found that 
two destabilization points exist for the DBB configuration one residing in the AF phase and one 
deep in the PO phase.
Specifically for $q<0$ the single in this case negative energy mode 
appearing in the BdG of Fig.~\ref{Figure5}(d) decreases in frequency and
crosses the spectral plane at $q_{cr}\approx-0.004$ rendering these entities unstable for this 
value of $q$ and thereafter. 
A result that is further supported by the finite growth rate, ${\rm Im}(\omega)\neq 0$, shown 
for these negative QZ energies in Fig.~\ref{Figure5}(e).
This destabilization is related to a composite motion of the DBB 
structure in the parabolic trap that we will soon trace in the dynamics. 
Contrary to the above destabilization yet another critical point occurs for the DBB solution for 
positive values of $q$. 
The latter appears at $q_{cr}\approx0.994$, i.e., the end point of the loop bifurcation illustrated 
in Fig.~\ref{Figure5}(e) above which DBB solitons cease to exist giving their place to a single 
dark solitary wave occupying the zeroth spin sublevel. 
This observation along with the second destabilization of the AM in this PO regime 
suggest the presence of a pitchfork bifurcation.  
In order to infer the existence of the latter we performed the corresponding stability 
analysis of the PO dark states (results not shown here for brevity). 
Interestingly enough, it is found that, even though dark solitons exist for all QZ energies in $q\in [0, 1.5]$, 
a narrow instability interval occurs at $q\in [0.994, 1]$ for these stationary states. 
Within this $q$ interval also the Krein signature changes sign from negative before the lower 
bound to positive after the upper bound. 
This, in turn, means that the dark soliton destabilizes slightly below unity and restabilizes for $q>1$. 
It is in this interval that indeed the above identified DBB solitons coexist with the single 
dark ones in a {\it subcritical} pitchfork bifurcation. 
Namely, dark solitons exist as stable configurations, for $q<0.994$, while their DBB counterparts are unstable. 
The collision of the two (and associated disappearance of the bright component of the DBB's) destabilizes 
the darks for  $q\in [0.994, 1]$, while for $q>1$, the relevant real eigenvalue pair returns to the imaginary axis, 
restabilizing the relevant dark state.

Direct evolution of the above-identified configurations slightly below and above 
the phase transition threshold at $q=0$ reveals that the DBB solitons undergo in both cases an overall breathing motion. 
Notice, for instance, the multi-frequency evolution of the DBB stationary state when excited along its 
most unstable eigendirection [Fig.~\ref{Figure5}(f)-(h)], entailing also an irregular population transfer 
between the spin-components [Fig.~\ref{Figure5}(i)]. 
These features are absent for $q<0$ (results not shown). 
In sharp contrast to the above dynamics for well-defined DBB solitons, i.e. away from the transition 
and the critical point, a well-defined, in-trap oscillation of the perturbed DBB wave is observed 
[Fig.~\ref{Figure5}(j)-(l)] for evolution times up to $t=5\times 10^{3}$ with the respective populations, 
$n_{m_F}(t)$, remaining constant for all times [Fig.~\ref{Figure5}(m)].  

\subsection{Easy-Axis symmetry broken DDB solitons}

Moving on to the EA phase, i.e., for $c_{1}=-5\times 10^{-3}$ and $q<0$,
again DDB stationary states are successfully identified.
However, contrary to the DDB solutions found in the AF phase here 
the DDB waves exhibit unequally populated $m_{F}=\pm 1$ spin-components
as can be seen in the insets of Fig.~\ref{Figure9}(a) and also in the relevant populations of 
Fig.~\ref{Figure9}(b).   
\begin{figure*}[htb]
 \centering \includegraphics[width=0.88\textwidth]{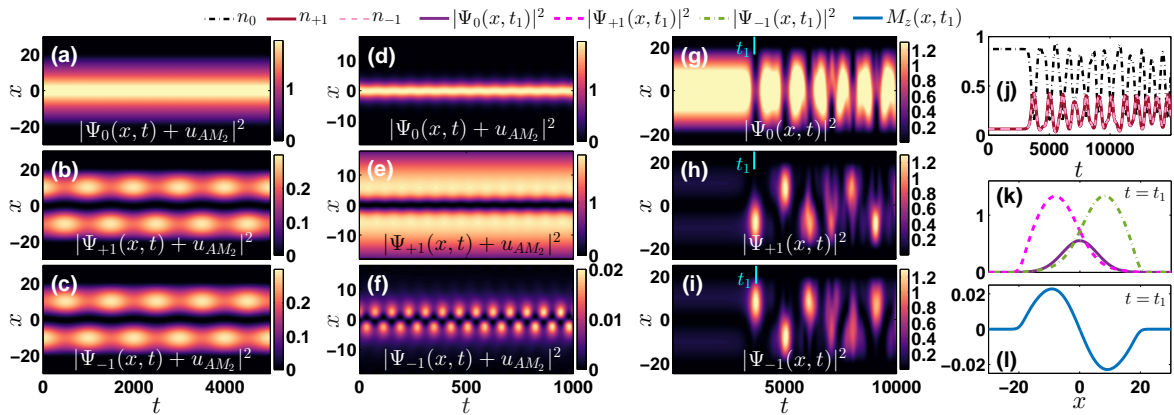}
  \caption{(a)-(f) [(g)-(i)] Spatio-temporal evolution of the density, $|\Psi_{m_{F}}+u_{AM_2}|^2$ [$|\Psi_{m_{F}}(x,t)|^2$], 
  of a perturbed with $u_{AM_2}$ [unperturbed but deformed] EA DDB soliton. 
The selected values of $q$ are $q=-0.002$, $q=-0.08$ [$q=0.002$], i.e. lying respectively in the 
region where the second and the fourth [fifth] bifurcation shown in Fig.~\ref{Figure9}(e) appear. 
(j) Evolution of the populations, $n_{m_F}(t)$, corresponding to the deformed configuration. 
(k) Profile snapshots of the deformed densities and (l) the associated to them local magnetization, $M_z(x)$, at $t_1=3690$. 
In all cases $m_{F}=0,\pm 1$ while $\Omega=0.1$, $\mu_{0,\pm1}=2$, $c_{1}=-5\times 10^{-3}$, and $c_{0}=1$. 
\label{Figure13}} 
\end{figure*}
Specifically, for these states the dark soliton of e.g. the $m_F=-1$ spin state is suppressed 
for most of the $q$ values within the region of existence, i.e. $q\in (-0.515, 0.007)$, of this configuration. 
Notably, such waves preserve the symmetry (i.e., equal population) of the $m_{F}=\pm1$ components 
for $q\in [-0.009, 0.007)$, namely including also the transition point ($q=0$) that separates the EA and the EP phases. 
For the remaining QZ energies lying in the aforementioned $q$ interval the symmetry is partially preserved, 
i.e. the $m_F=-1$ is still populated. 
This result is encoded in the magnetization and the polarization properties of the DDB solutions 
which assume values $0<M_z<1$ and $-1<P<1$ respectively reflecting the non-negligible population of all three spin-components. 
More precisely, starting with $P=1$ ($M_z=0$) for $q>0.007$, the relevant quantity decreases (increases) 
when moving towards $q<0$ and approaches the value of $P=-1$ ($M_z=+1$) for $q<-1$, 
i.e. deep in the EA phase [Fig.~\ref{Figure9}(a)]. 
These symmetric DDB solitons are fundamentally different (structurally) than the relevant GS 
configuration in this parametric regime. 
The latter, according to the phase diagram of Fig.~\ref{Figure1}(a), favors symmetry broken 
states that are fully magnetized along the $+z$- or $-z$-spin direction, i.e. configurations 
that have either the $m_F=+1$ or the $m_F=-1$ component solely populated. 
As such, for $q\in (-0.515,-0.2]$ the dark soliton of the $m_{F}=-1$ spin state becomes narrower, 
in an almost exponentially decaying manner, as can be seen from the behavior of its width, 
$w_{-1}(q)$, shown in Fig.~\ref{Figure9}(c). 
In particular, around $q\approx -0.515$ the configuration is deformed to a symmetry broken almost 
fully magnetized ($M_{z}\approx+1$) DB soliton that exists for $q\in[-1.5, -0.515]$ occupying the $m_F=+1$ and $m_F=0$ spin 
components (similarly, of course, there is a state occupying the $m_F=-1$ and $m_F=0$ states). 
Additionally, as the transition point is approached from below, $q\rightarrow 0^{-}$, also the population, $n_0(q)$, 
and width, $w_0(q)$, of the bright component increases and the DDB solitons deform even further. 
Specifically, for $q\in (-0.004, 0.007)$ a structure with a bright component that overfills the dark 
wells while gradually morphing into a TF profile can be identified, as shown in the upper left inset of Fig.~\ref{Figure9}(a). 
This deformed DDB structure enters the EP phase, which favors all three $m_F$ components to be simultaneously occupied, 
but already at $q\approx 0.008$ the unmagnetized GS of the PO phase is reached.  

The BdG analysis of the above-discussed soliton solutions illustrated in Fig.~\ref{Figure9}(d), (e) 
reveals that DDB solitons possess potentially unstable eigendirections (although they also possess stability intervals). 
This result can be inferred by the finite imaginary eigenfrequencies (or instability growth rates), ${\rm Im}(\omega)$, 
shown in Fig.~\ref{Figure9}(e). 
The relevant unstable $q$ intervals for the DDB wave are $[-0.384, -0.329]$, $[-0.111, -0.066]$, $[-0.052, -0.022]$, 
$[-0.042, -0.003]$ and $[-0.016, 0]$ respectively. 
Closely inspecting the relevant ``gaps'' in the trajectory of each of the two 
AMs depicted in Fig.~\ref{Figure9}(d) it becomes apparent that the first (from negative to positive QZ values) 
loop bifurcation shown in Fig.~\ref{Figure9}(e) is associated with the lower-lying AM. 
Consecutively the second loop is related to the higher-lying AM and so on for 
the remaining three instability bubbles. 
Notice that the last bifurcation possesses also the larger instability growth rate that stems 
from an eigenfrequency zero crossing of both the higher- and the lower-lying AM appearing at $q_{cr}=0$. 
For $-0.515<q<-0.384$ the state remains linearly stable having, however, a minuscule $m_F=-1$ component. 
For more negative values of $q$, the fully magnetized linearly stable DB solitons are present in the spin-1 system. 
To confirm the above stability analysis findings we have monitored the dynamical evolution of the DDB solutions 
in all of the above-identified instability intervals and our results can be summarized as follows. 
Among the two modes that appear in the BdG spectrum of Fig.~\ref{Figure9}(d) the lower one is related to the weak 
amplitude in trap oscillation of the DDB wave. 
The higher mode is responsible for the larger in amplitude anti-phase oscillation of the involved dark solitons. 
Additionally, it turns out that even when these states are found to be dynamically unstable they have 
remarkably long lifetimes that support their experimental observation in existing spinor settings~\cite{Bersano}.
\begin{figure*}[htb]
 \centering \includegraphics[width=0.8\textwidth]{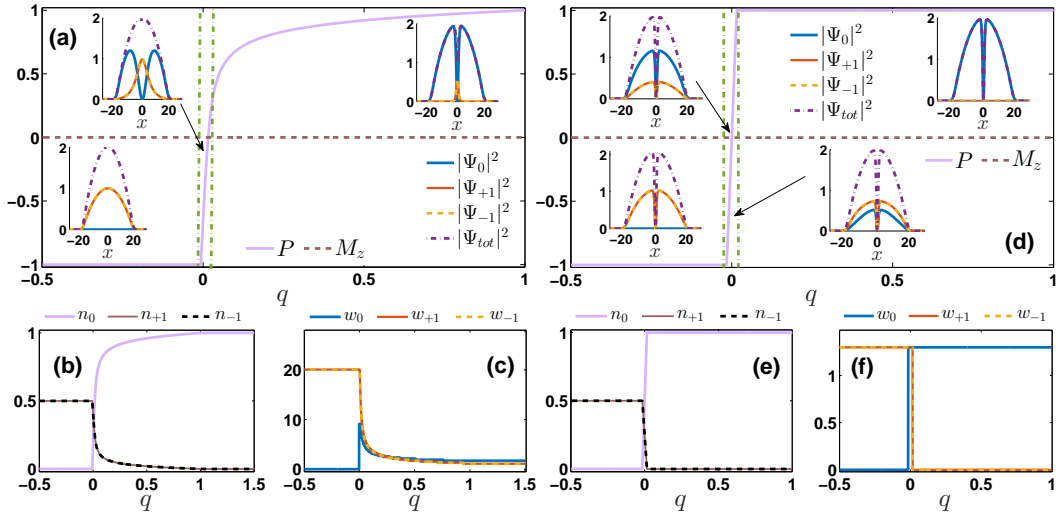}
 \caption{(a) [(d)] Polarization $P$, and magnetization, $M_z$, of a DBB [DDD] wave within the EP phase upon 
 varying $q$ towards the EA and PO phases. Insets from bottom left and on (see also black arrows) illustrate 
 characteristic soliton profiles for $q=-0.1$ [$q=-0.1$], $q=-0.02$ [$q=-0.008$] and $q=0.1$ [$q=0.003$ and $q=0.5$]. 
 (b), (e) Populations, $n_{m_{F}}$, and (c), (f) soliton widths, $w_{m_{F}}$, with $m_F=0,\pm1$, of a DBB and 
 a DDD soliton state respectively, upon varying $q$. 
 In (a) and (d) vertical dashed-dotted (green) lines denote the boundaries of the EP phase. 
 Other parameters used are $\Omega=0.1$, $\mu_{0,\pm1}=2$, $c_{1}=-5\times 10^{-3}$, and $c_{0}=1$.
\label{Figure15}} 
 \end{figure*}
A case example showcasing the particle-like oscillations that a perturbed DDB stationary state undergoes 
is presented for $q=-0.002$ in Fig.~\ref{Figure13}(a)-(c). 
Notice that indeed the DDB wave remains intact for all times up to $t=5\times 10^3\approx 4.55$s (in dimensional units). 
More specifically, by perturbing the DDB soliton with the eigenvector associated with the first AM leads to an 
oscillation of the wave within the trap (results not shown for brevity). 
On the other hand, the second mode results in the formation of two atomic blobs to which the dark solitons split 
the entire condensate [Fig.~\ref{Figure13}(a)-(c)]. 
These blobs execute an anti-phase oscillation alternating across the two dark components 
and across the two sides (left and right) of the dark solitary wave in each component. 
It is important to note that this type of periodic orbits, such as the ones emerging here, 
is a natural by-product of the AM-induced instabilities and the corresponding Hamiltonian-Hopf bifurcations 
(which are well-known in dynamical systems to generate --or potentially destroy-- such periodic orbits). 
This latter anti-phase oscillation becomes even more pronounced especially for the $m_F=-1$ component as $q$ 
decreases further towards the formation of DB solitons that occupy the $m_F=0$ and $m_F=+1$ magnetic 
sublevels [Fig.~\ref{Figure13}(d)-(f)]. 
Evidently, as $q$ decreases further and e.g. for $q=-0.08$ illustrated in Fig.~\ref{Figure13}(d)-(f), 
the population of the $m_F=-1$ magnetic sublevel becomes significantly suppressed.
 
Contrary to the above-described dynamics, the picture is drastically altered when considering the 
deformed DDB stationary states that exist near the transition point ($q=0$). 
For instance here, by monitoring $|\Psi_{m_F}(x,t)|^2$ for QZ energy shifts that lie within the last 
bifurcation [Fig.~\ref{Figure9}(e)] reveals that these transient states for evolution times of 
the order of $t\approx 4500$ destabilize towards states that consist of Gaussian-like (localized) 
structures hosted in the $m_F=0$ component. 
These localized density blobs are not of permanent character as is evident in Fig.~\ref{Figure13}(g) 
but they revive in an almost periodic manner. 
In every recurrence event, the corresponding symmetric spin-components bear droplet-like configurations 
that appear in an alternating fashion either in the $m_{F}=+1$ or in the $m_{F}=-1$ component [Fig.~\ref{Figure13}(h), (i)]. 
This behavior essentially reflects the continuous spin transfer between the $m_F=0$ and $m_F=\pm1$ 
taking place during evolution [Fig.~\ref{Figure13}(j)]. 
Inspecting the density profiles of the evolved states [Fig.~\ref{Figure13}(k)] unveils that 
the wavefunction of the zeroth magnetic sublevel acts as a repulsive barrier pushing outwards, 
with respect to the trap center, the symmetric spin-components that develop in between them a DW~\cite{Ektor_DW}. 
Measuring the local magnetization, $M_z(x)$, e.g. at $t_1=3690$ where this dynamically formed state emerges 
for the first time [Fig.~\ref{Figure13}(l)], reveals that such a configuration bears indeed a DW character 
across which $M_z(x)$ changes sign~\cite{Ektor_DW}. 
Such a magnetic entity holds close similarities to the so-called magnon drop, namely a soliton-like 
object that has the direction of magnetization in each core opposite to its surroundings~\cite{Macia,Divinskiy}.

\subsection{Nematic DBB and DDD solitons}   
\begin{figure*}[htb]
 \centering \includegraphics[width=0.9\textwidth]{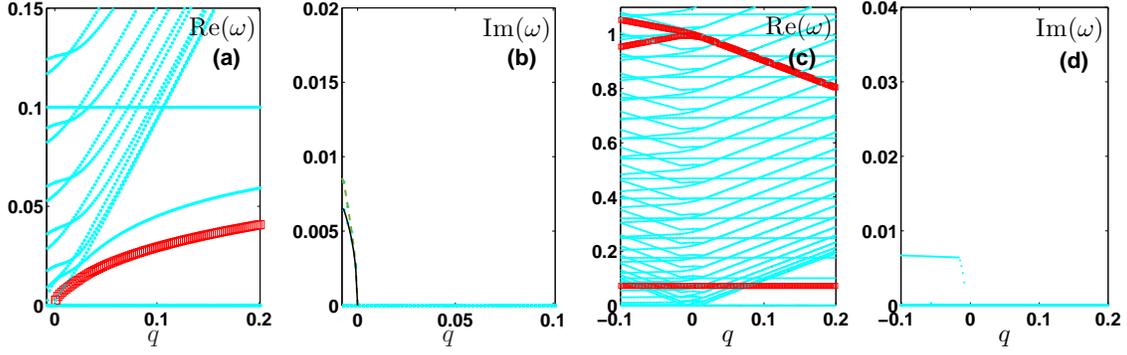}
 \caption{BdG spectrum of DBB [DDD] waves existing in the EP phase as we vary $q$ towards both the EA and the PO phases. 
 (a) [(c)] Real part, ${\rm Re}(\omega)$, of the corresponding eigenfrequencies as a function of the QZ coefficient $q$. 
 The trajectory of the single [three] AM [AMs] present in this spectrum is shown by red squares (see text). 
 (b) [(d)] Imaginary part, ${\rm Im}(\omega)$, of the respective eigenfrequencies. 
 The eigenvalue zero crossing for the DBB waves occurs at $q_{cr}=0$ leading to ${\rm Im}(\omega)\neq 0$ 
 while solid (black) and dashed (green) lines in (b) are used as a guide to the eye. 
 Other parameters used are $\Omega=0.1$, $\mu_{0,\pm1}=2$, $c_{1}=-5\times 10^{-3}$ and $c_{0}=1$.
 \label{Figure17}}
 \end{figure*}

Subsequently we study the properties of nonlinear structures in the EP phase. 
The latter as per the phase diagram of Fig.~\ref{Figure1}(b) corresponds to $c_1=-5\times 10^{-3}$ 
and $0<q<q_{th}$ supporting both DBB and DDD stationary states. 
These distinct nonlinear excitations illustrated respectively in the insets of Fig.~\ref{Figure15}(a) 
and ~\ref{Figure15}(d), appear to be unmagnetized since $M_z=0$ in both cases while having a 
nontrivial polarization as $q$ is varied. 
Moreover, the DBB entities are found to be significantly broader around $q=0$ when compared 
to the highly localized DDD solitons [see top left insets in Fig.~\ref{Figure15}(a) and ~\ref{Figure15}(d)]. 
Interestingly, DBB solitons deform rapidly, i.e. soon after the transition point separating the EP 
to EA phases is crossed and for $q\approx-0.007$, into the metastable state of the EA phase that has 
equally populated symmetric components [Fig.~\ref{Figure15}(b)] when compared to the slower, around $q\approx-0.016$, 
deformation of the DDD solitons into two dark ones equally populating the $m_{F}=\pm1$ spin states [Fig.~\ref{Figure15}(e)]. 
Notice that in the former DBB case, the dark soliton in the $m_{F}=0$ component has disappeared 
and only a Thomas-Fermi type profile remains in the $m_{F}=\pm 1$ components. 
Importantly though, as $q$ is increased so as to approach the critical point $q=q_{th}$ that separates 
the EP and the PO phases, a rather sharp transitioning takes place for DDD solitons when compared to 
the significantly smoother one exhibited by the DBB stationary states. 
This sharp versus smooth transition can be inferred by inspecting the relevant slopes of the polarization for $0<q<q_{th}$. 
Specifically, it is found that DDD solitons morph faster, i.e. for $q=0.016$, into a single dark state 
occupying the $m_F=0$ component. 
This observation is in agreement with the prediction from the GS analysis threshold value of the quadratic energy term, 
which in turn suggests that for $q=q_{th}$, the PO phase should be reached [top right inset in Fig.~\ref{Figure15}(d) 
and Fig.~\ref{Figure15}(e), (f)]. 
Contrary to this deformation, it is only around $q=1$ that the polarization measured for DBB solitons 
asymptotes to $P=1$ [Fig.~\ref{Figure15}(a)]. 
The latter together with the relevant negligible populations, $n_{m_F}(q)$ [Fig.~\ref{Figure15}(b)], 
and widths, $w_{m_F}(q)$ [Fig.~\ref{Figure15}(c)], of the bright matter waves hosted in the symmetric $m_F=\pm1$ 
spin states designates the transition towards a single dark state existing in the PO phase. 
From the above analysis we can conclude that the boundary separating the EP and the PO phases can be less 
transparent when considering nonlinear excitations instead of ground states. 
This is especially so for states like the DBBs for which the two components play a complementary 
role, i.e., the bright solitary waves in the $m_{F}=\pm 1$ components fill the hole generated by the dark 
one in the $m_{F}=0$ component.

Our BdG results reveal that both DBB and DDD solitons are stable 
configurations within the EP phase as can be inferred by the zero imaginary 
part shown in Fig.~\ref{Figure17}(b) and Fig.~\ref{Figure17}(d) respectively. 
Also stable are the single dark solitons (into which the above DBBs and DDDs morph) in the PO phase, 
whose stability analysis simply leads to the standard oscillatory motion, with oscillation 
frequency $\omega_{osc}=\Omega/\sqrt{2}$ in the TF regime, known for harmonically trapped dark solitons~\cite{Frantz_dark}. 
Furthermore the existence of a single and three AMs pertaining to the DBB and the DDD configuration 
respectively can also be seen in the relevant real part of the spectrum illustrated in Fig.~\ref{Figure17}(a) 
and Fig.~\ref{Figure17}(c). 
Once again, it appears that the number of components bearing a dark solitary wave determines 
the number of AMs within the state of interest.
\begin{figure*}[htb]
 \centering \includegraphics[width=0.8\textwidth]{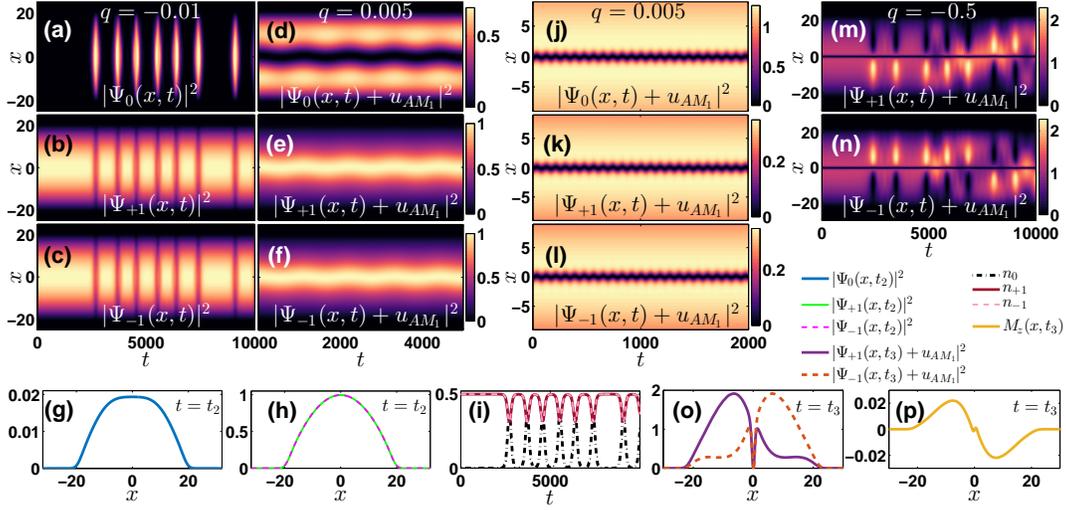}
 \caption{(a)-(c) [(d)-(f)] Spatio-temporal evolution of the density, $|\Psi_{m_{F}}(x,t)|^2$ [$|\Psi_{m_{F}}+u_{AM_1}|^2$], 
 of a metastable [DBB] state occurring for $q$ in the EA [EP] phase. 
 (g), (h) Profile snapshots of the densities of the above metastable states at $t_2=3030$ to illustrate the droplet 
 formation and (i) temporal evolution of the ensuing populations, $n_{m_F}(t)$. 
 (j)-(l) [(m), (n)] Same as (d)-(f) for an EP DDD [DD] soliton [but for $q$ in the EA phase]. 
 (o) Density profiles of the symmetric spin sublevels and (p) of the spatial magnetization, $M_z(x)$, at $t_3=2500$ 
 when the emergent spin-wave is spontaneously nucleated (see legends). 
 In all cases $m_{F}=0,\pm 1$ while from left to right the distinct columns correspond to $q=-0.01$, $q=0.005$ and $q=-0.5$ 
 respectively. The remaining system parameters are $\Omega=0.1$, $\mu_{0,\pm1}=2$, $c_{1}=-5\times 10^{-3}$ and $c_{0}=1$.
\label{Figure19}} 
 \end{figure*}
However, and as far as the DBB solutions are concerned, as $q$ decreases so as to enter the EA phase 
our stability analysis shows that an eigenfrequency zero crossing occurs right at 
the transition point ($q=0$) suggesting the destabilization of the DBB wave. 
Below this point and specifically for $q<-0.007$ different types of stationary states exist for the spin-1 system. 
These new metastable states consist of an unpopulated $m_F=0$ component and two nearly TF density profiles 
occurring in the other two equally populated symmetric magnetic sublevels. 
The finite growth rate, $\rm{Im}(\omega)$, depicted in Fig.~\ref{Figure17}(b) unveils the emergence 
of these new unstable configurations. 

Turning to the DDD solitons, for these negative QZ energies, we can easily deduce that also these waves gradually deform. 
Their destabilization as detected by the finite growth rate observed in Fig.~\ref{Figure17}(d) occurs at $q\approx -0.009$ 
rendering also these DDD solitons unstable for $q\in [-0.016, -0.009]$. 
However, since $\rm{Im}(\omega)\neq 0$ even deeper in the EA phase this further implies 
that also the DD solitons that are formed for $q<-0.017$ exist as unstable configurations 
for this value of $q$ onward within the EA phase. 
Interestingly, the instability of these states is caused by an imaginary eigenfrequency reflecting the co-existence 
of these two components. 
	
Confirmation of the above-obtained stability analysis results is provided in Fig.~\ref{Figure19}(a)-(i) 
for the DBB solutions and in Fig.~\ref{Figure19}(j)-(p) for the DDD waves. 
Notice the coherent particle-like oscillations observed for the stable DBB [Fig.~\ref{Figure19}(d)-(f)] 
and DDD [Fig.~\ref{Figure19}(j)-(l)] solitons for $q>0$ when compared to the unstable evolution of the densities for $q<0$. 
The spatio-temporal evolution of the metastable states depicted in Fig.~\ref{Figure19}(a)-(c) is apparently rather similar 
to the one found for the relevant states upon crossing the EA to EP phase boundary [see Fig.~\ref{Figure13}(g)-(i)]. 
Here, however, the two symmetric nonzero $m_F$ components lose atoms towards the $m_F=0$ state 
in a nearly periodic fashion as a result of the instability. 
This dynamical evolution leads to states featuring a flat-top, droplet-like profile [Fig.~\ref{Figure19}(g)] 
and nearly TF wavefunctions occupying, respectively, the $m_F=0$ and $m_F=\pm1$ components [Fig.~\ref{Figure19}(h)]. 
Coherent population transfer accompanies the periodic revival of these states [Fig.~\ref{Figure19}(i)] 
which remain nematic, i.e. having zero magnetization, during evolution. 
This way, they preserve the magnetic properties expected for an EP configuration. 
Furthermore, these unmagnetized structures appear to robustly re-emerge up to times $t>8\times 10^{3}$ 
(when this apparent periodicity is modified).

Next we monitor the relevant unstable evolution of the perturbed DD waves. 
Strikingly, their dynamics entails completely new features as shown in Fig.~\ref{Figure19}(m), (n). 
In this case, on top of the perturbed DD solitons, localized states having widths significantly 
larger than the healing length, which is the characteristic length scale of regular solitons, develop. 
These localized matter waves consist of density humps followed by density dips building on top of the BEC background. 
They further emerge in an alternating fashion not only within but also between the symmetric spin-components 
[see the density profiles at $t_3=2500$ shown in Fig.~\ref{Figure19}(o)]. 
These structures are reminiscent of phase separated states that have been widely considered in 
multi-component condensates~\cite{Stringari,siambook}. 
This is also reflected in the antisymmetric extended spatial profile of $M_z$ as can be seen in Fig.~\ref{Figure19}(p). 
This quantity reflects the distinct spin domains formed among the $m_F=\pm1$ components, while the dark soliton 
remains in the middle being shared by the two otherwise phase-separated $m_F=\pm 1$ components. 

\begin{figure*}[htb]
 \centering \includegraphics[width=0.8\textwidth]{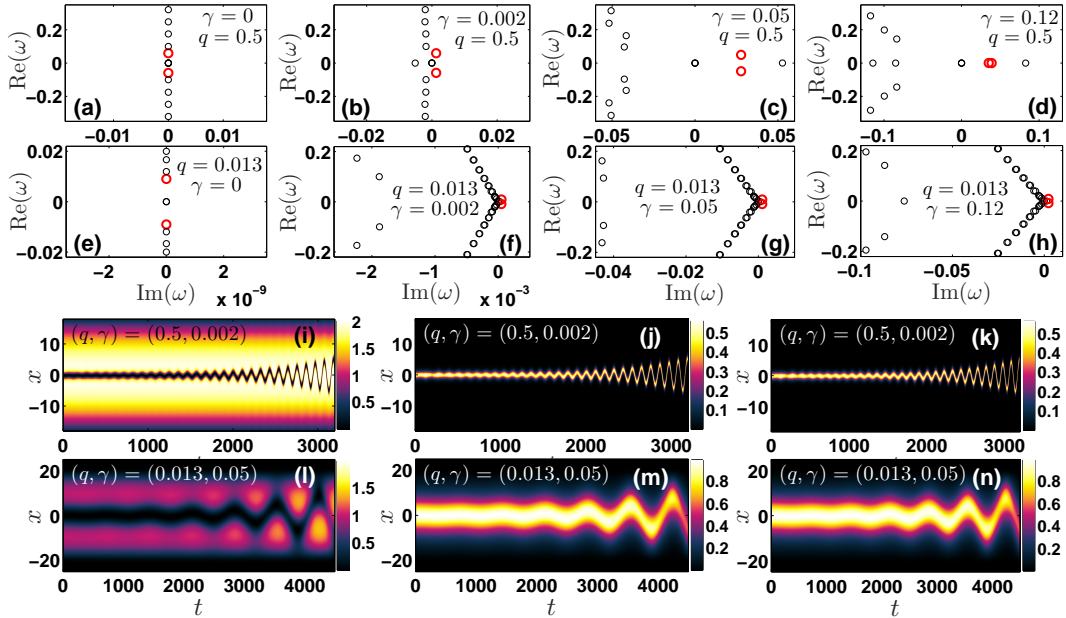} 
 \caption{(a)-(d) [(e)-(h)] Spectral plane of a stationary PO [EP] DBB soliton 
 for $c_{1}=5\times 10^{-3}$ [$c_{1}=-5\times 10^{-3}$] for different values of $(q, \gamma)$ (see legends).
 Red circles denote the anomalous eigenfrequencies which lead to a Hopf bifurcation and an oscillatory instability of the DBB wave. 
 (i)-(k) [(l)-(n)] Spatio-temporal evolution of the density, $|\Psi_{m_{F}}(x,t)+u_{AM_1}|^2$, of a perturbed PO [EP] DBB 
 soliton for $(q, \gamma)=(0.5, 0.002)$ [$(q, \gamma)=(0.013, 0.05)$]. The remaining system parameters are $\Omega=0.1$, 
 $\mu_{0,\pm1}=2$, and $c_{0}=1$. 
 \label{FigureFT}}
 \end{figure*}

\section{Finite temperature effects on spinor solitons}\label{FT}

We now consider the case where the spinor gas is exposed to finite temperatures. 
In order to account for the latter we extend previous considerations pertinent to single-~\cite{PGK_darkFT} 
and two-component BECs~\cite{PGK_DBFT} to the spinorial case at hand. 
In particular, by assuming that only the thermal modes along the axial $x$-direction are occupied, 
we utilize the following system of three coupled dissipative GPEs --so-called DGPEs--
\begin{eqnarray}
\left(i-\gamma\right) \partial_t \Psi_{0}&=&
\Big[\mathcal{\tilde{H}}_{0} 
+ c_{0} \left( |\Psi_{+1}|^2 + |\Psi_{0}|^2 + |\Psi_{-1}|^2 \right)\Big]\Psi_{0}\nonumber\\
&+&c_{1}\left( |\Psi_{+1}|^2 + |\Psi_{-1}|^2 \right)\Psi_{0} \nonumber\\
&+& 2c_{1}\Psi_{+1}\Psi^{*}_{0} \Psi_{-1}, 
\label{eq:0_FT} \\
\left(i-\gamma\right) \partial_t \Psi_{\pm 1} &=&
\Big[\mathcal{\tilde{H}}_{0} 
+ c_{0} \left( |\Psi_{+1}|^2 + |\Psi_{0}|^2 + |\Psi_{-1}|^2 \right)\Big]\Psi_{\pm 1}\nonumber\\
&+&c_{1}\left( |\Psi_{\pm1}|^2 + |\Psi_{0}|^2 - |\Psi_{\mp 1}|^2 \right)\Psi_{\pm 1}\nonumber\\
&+&q\Psi_{\pm 1}+c_{1}\Psi^{*}_{\mp 1}\Psi^{2}_{0}.
\label{eq:pm1_FT} 
\end{eqnarray}
In Eqs.~(\ref{eq:0_FT})-(\ref{eq:pm1_FT}) $\mathcal{\tilde{H}}_{0} \equiv \mathcal{H}_{0}-\mu_{m_F}$, 
while the dimensionless parameter $\gamma_{0}=\gamma_{+1}=\gamma_{-1}\equiv \gamma$, is associated 
with the system's temperature~\cite{PGK_vortex}. 
Particularly, $\gamma \ll 1$ lying in the range of $2 \times 10^{-4}-2\times 10^{-3}$ for temperatures 
(in dimensional units) of the order of $10-100$nK~\cite{PGK_vortex}. 
Before appreciating the effect of the damping term $\gamma$ on the statics as well as the dynamics 
of the spinor solitons identified herein, we note that contrary to the Hamiltonian, $\gamma=0$, 
case the negative energy eigenmodes are expected for $\gamma \neq 0$ to bifurcate towards the 
right half-plane of the excitation spectrum~\cite{PGK_darkFT,Kapitula}. 
Additionally, the corresponding positive energy eigenmodes will move on the left half-plane 
in this DGPE setting~\cite{PGK_darkFT,PGK_DBFT}. 
The above-described spectral displacement implies in turn an immediate dynamical instability 
of all the (excited state) spinorial entities discussed herein. 

An example of such a migration of the involved eigenmodes is depicted in Fig.~\ref{FigureFT}(a)-(h), 
for a PO (top row) and an EP (second row) DBB soliton for $q=0.5$, $c_{1}=5\times 10^{-3}$ 
and $q=0.013$, $c_{1}=-5\times 10^{-3}$ respectively, under a $\gamma=0-0.12$ variation. 
Evidently, as $\gamma$ increases the eigenfrequency pair previously associated with negative 
Krein signature (denoted by red circles), moves to the right half-plane acquiring progressively a decreasing real part. 
This behavior continues until the eigenfrequency pair collides and subsequently splits along the imaginary 
eigenfrequency axis, giving in turn rise to a purely exponential instability. 
The anti-damping, i.e. oscillation of growing amplitude, of both the PO [Fig.~\ref{FigureFT}(i)-(k)] 
and the EP [Fig.~\ref{FigureFT}(l)-(n)] DBB solitons when $\gamma\neq 0$ can be directly contrasted with their 
respective constant amplitude in-trap oscillation for $\gamma=0$ [Fig.~\ref{Figure5}(j)-(l) 
and Fig.~\ref{Figure19}(d)-(f) respectively]. 
This anti-damping is weaker when the bright soliton component ``filling" of the dark notch is more pronounced~\cite{PGK_DBFT} 
as is the case of nematic DBB configurations  [see Fig.~\ref{FigureFT}(l)-(n) and the top left inset of Fig.~\ref{Figure15}(a)]. 
Finally, analogous dynamical results are observed for all of the remaining spinorial entities (results not shown).

\section{Conclusions}\label{conclusions}

The complete phase diagram of solitonic nonlinear excitations that arise in the distinct phases of ferromagnetic 
and antiferromagnetic 1D spin-1 harmonically trapped BECs, being unprecedented thus far, 
has been extracted and explored in detail. 
In particular, spinor matter-waves in the form of DDD DDB and DBB solitons, are tackled 
in the spin-QZ energy-plane, $(c_1,q)$, being further distinguished and classified 
in terms of their magnetic, stability and dynamical properties. 
This effort has been strongly motivated by recent experiments focused on studying 
the magnetic or not soliton excitations forming in spin-1 Bose gases~\cite{Bersano,Farolfi,Chai,Lannig,raman2}. 
Specifically, it is found that DDB solitons exist in the antiferromagnetic and the easy-axis phases, 
being unmagnetized and unstable configurations in the former and magnetized, experiencing also stable intervals in the latter phase. 
Unmagnetized DBB solitons are identified in the easy-plane and the polar phase as stable and unstable entities respectively, 
while the coexistence of easy-plane DBB solitons with stable and nematic DDD ones is showcased. 
Remarkably, all of the above-mentioned stable and unstable waveforms, whose dynamics entails predominantly 
particle-like translational or breathing oscillations, experience lifetimes ranging from one two several seconds, 
corroborating their direct experimental relevance and potential observability. 
Alterations of the statics and dynamics of all of these spinors when exposed to finite temperatures 
have also been studied. 
Here, the anti-damping in trap oscillation of all states is unravelled, 
being progressively suppressed for larger bright soliton component ``fillings'' 
of the dark notch, generalizing this way earlier findings to the spin-1 setting. 

Focusing on the relevant deformations of each principal spinor soliton far from and around the associated 
transition threshold it is demonstrated that antiferromagnetic DDB states deep in the antiferromagnetic 
phase morph into a symmetric DD configuration while immediately after crossing the transition boundary 
are abruptly deformed in the ground state of the polar phase. 
Three distinct deformations occur for easy-axis DDB waves, namely from fully magnetized stable DB solitons 
deep in the easy-axis, to metastable states near the easy-axis--easy-plane threshold and 
finally to the ground state of the polar phase. 
Interestingly, among these morphings, the metastable states develop into long-lived magnetic spin 
configurations that resemble the so-called magnon-drops~\cite{Macia,Divinskiy} 
with a characteristic domain-wall~\cite{Ektor_DW} building between the droplets and being imprinted in the local magnetization. 
Also polar and easy-plane DBB solitons deform with the former penetrating the antiferromagnetic phase leading 
to coexisting DDB and DBB waves. 
The polar DBB solitons feature two deformations: they abruptly morph either to the antiferromagnetic ground 
state or into stable single dark solitons deep in the polar regime. 
On the other hand, nematic DBB solitons turn into metastable states as the easy-plane--easy-axis 
threshold is crossed which evolve into nematic this time yet long-lived droplets. 
Highly localized DBB solitons occur for an easy-plane--polar transition before the final 
morphing of these states to single and linearly stable darks. 
Finally, nematic DDD solitons of the easy-plane phase experience an abrupt deformation 
to a single dark soliton deep in the polar regime while they gradually morph, when entering the easy-axis phase, 
into unstable magnetized symmetric DD configurations. 
Strikingly, these DD entities evolve into composite spin objects containing a central dark soliton and a spin-wave. 
They have finite local magnetization and remarkably long lifetimes. 
Evidently, a plethora of new entities are identified in this spin-1 setting, 
whose magnetic imprint can be probed experimentally.

Several extensions of the present work can be put forth. 
As a first step one can unravel the fate of the identified spin-1 soliton solutions 
subjected to quenches across the first and second order phase transition boundaries. 
Yet another interesting perspective would be to study interactions~\cite{Lannig,Meng,Szankowski} 
between the spinor solitons identified within each phase of the above-obtained phase diagram 
or even unravel lattices consisting of multiple spinorial solitons in analogy to the two-component 
settings e.g. of Refs.~\cite{Alejandro,Tsitoura}. 
In some cases where different solutions co-exist (e.g. the DBB and the DDD in the easy-plane phase), 
one could even consider collisions between different types of entities. 
Another aspect that the present work motivates further concerns the study of domain-wall 
configurations in suitable regimes of the relevant phase diagram (e.g., within the easy-axis phase). 
In the present setting we did not consider the role of three-body losses, motivated by the evident 
absence of their consideration in the array of recent experiments~\cite{Bersano,Farolfi,Chai,Lannig,raman2}. 
Yet, for longer times, such effects should naturally come into play and are worthwhile of separate consideration. 
Finally, generalizing the phase diagram of nonlinear excitations extracted herein in higher dimensions 
where vortex-bright states~\cite{tuckerman,Mukherjee} are expected to form would be also a fruitful future direction.

\section{Acknowledgements} 
P.G.K. is grateful to the Leverhulme Trust and to the Alexander von Humboldt Foundation for support
and to the Mathematical Institute of the University of Oxford for its hospitality. 
G.C.K and P.S. gratefully acknowledge financial support by the Deutsche Forschungsgemeinschaft (DFG) 
in the framework of the SFB 925 ``Light induced dynamics and control of correlated quantum systems". 
S.I.M. gratefully acknowledges financial support in the
framework of the Lenz-Ising Award of the University of Hamburg.

{}

\begin{thebibliography}{60}
\bibitem{FWHM} Note that in order to estimate the solitonic widths we measure the 
full-width-at-half-maximum (FWHM) of the ensuing matter waves for varying $q$.
\end{thebibliography}
\end{document}